\documentclass{aastex62}

\submitjournal{ApJ}

\shorttitle{Buildup of Abiotic Oxygen and Ozone}
\shortauthors{Kleinb{\"o}hl et al.}

\begin{document}

\title{Buildup of Abiotic Oxygen and Ozone in Moist Atmospheres of Temperate Terrestrial Exoplanets
and its Impact on the Spectral Fingerprint in Transit Observations}

\author{Armin Kleinb{\"o}hl}
\affil{Jet Propulsion Laboratory \\
California Institute of Technology \\
4800 Oak Grove Dr. \\
Pasadena, CA, 91109, USA.}

\author{Karen Willacy}
\affiliation{Jet Propulsion Laboratory \\
California Institute of Technology \\
4800 Oak Grove Dr. \\
Pasadena, CA, 91109, USA.}

\author{A. James Friedson}
\affiliation{Jet Propulsion Laboratory \\
California Institute of Technology \\
4800 Oak Grove Dr. \\
Pasadena, CA, 91109, USA.}

\author{Pin Chen}
\affiliation{Jet Propulsion Laboratory \\
California Institute of Technology \\
4800 Oak Grove Dr. \\
Pasadena, CA, 91109, USA.}

\author{Mark R. Swain}
\affiliation{Jet Propulsion Laboratory \\
California Institute of Technology \\
4800 Oak Grove Dr. \\
Pasadena, CA, 91109, USA.}



\begin{abstract}

We investigate the abiotic production of oxygen and its photochemical byproduct ozone
through water vapor photolysis in moist atmospheres of temperate terrestrial exoplanets.
The amount of water vapor available for photolysis in the middle atmosphere of a planet
can be limited by an atmospheric cold-trap, the formation of which largely depends on the amount of non-condensable gases.
We study this effect using a photochemical model coupled to a 1D radiative-convective equilibrium model
in atmospheres with N$_2$, CO$_2$ and H$_2$O as the main constituents.
We find that in atmospheres with a low N$_2$ inventory, water vapor mixing ratios
in the middle atmosphere can be over two orders of magnitude higher compared to atmospheres
with an Earth-like N$_2$ inventory. Without a strong surface sink, the non-condensable oxygen
can build up rapidly, drying out the upper atmosphere. With a moderate surface sink,
the planet can approach a steady state with significant oxygen mixing ratios in which oxygen production is balanced by surface uptake.
We use a radiative transfer model to study the spectroscopic fingerprint of these atmospheres
in transit observations. Spectral signatures of abiotic oxygen and ozone can be of comparable magnitude
as in spectra of Earth seen as an exoplanet. Middle atmospheric water vapor is unlikely to be a usable indicator
of the abiotic origin of oxygen because of the influence of oxygen on the water vapor distribution. This suggests that atmospheric oxygen
and ozone cannot be used as binary bioindicators and their interpretation will likely require atmospheric and planetary models.

\end{abstract}

\keywords{astrobiology --- planets and satellites: atmospheres --- planets and satellites: composition
--- planets and satellites: physical evolution --- planets and satellites: terrestrial planets --- radiative transfer}


\section{Introduction}
\label{intro}

The last two decades have seen a rapid increase in the detection and characterization of exoplanets.
Future missions like the Transiting Exoplanet Survey Satellite (TESS) are expected to find numerous
transiting exoplanets \citep{sullivan15}. A focus of these missions will be on the subset of transiting, terrestrial,
temperate exoplanets as they are the strongest candidates to harbor life as we know it.
Bioindicators in exoplanetary observations are features whose presence or abundance indicates
a biological origin. An important bioindicator for life as we know it is the existence of significant
amounts of atmospheric oxygen, which on Earth is largely formed through photosynthesis
and sustained at an atmospheric mixing ratio of 21\% \citep{desmarais02,meadows17}.
Oxygen has absorption features in the near-IR,
with the most prominent feature being the oxygen A-band at 760 nm (e.g. \citet{hitran12}).
In the mid-IR the best indicator of the presence of oxygen is its photochemical byproduct ozone,
which has a strong absorption band centered at 9.6 $\mu$m (e.g. \citet{desmarais02}).
Hence atmospheric oxygen and its photochemical byproduct ozone are commonly assumed
to be indicators of the presence of life.

An important question is under what conditions and by what processes oxygen and ozone can be produced by abiotically.
Constraining these processes and the conditions under which they occur will be essential to avoid
'false positive' detections of life, that is the interpretation of oxygen or ozone as a bioindicator
despite it being produced abiotically. \citet{meadows17} give a comprehensive overview on processes that
lead to the abiotic formation of oxygen identified to date. Several processes that lead to buildup of abiotic oxygen
rely on the photolysis of CO$_2$, first suggested by \citet{selsis02}.
In dense CO$_2$ atmospheres the photolysis of CO$_2$ can lead to O$_2$ mixing ratios of order 0.1\% \citep{hu12}.
The accumulation of oxygen from this process depends on the ratio of far-UV to near-UV incident radiation.
This ratio is often enhanced in M dwarf stars such that exoplanets orbiting in the habitable zones of M dwarfs
have the potential for higher abiotic oxygen amounts than Sun-like stars \citep{domagal14,tian14,harman15}.
On dessicated planets the lack of water inhibits the catalytic recombination of CO$_2$,
leading abiotic O$_2$ mixing ratios of order 15\% on planets orbiting
in the habitable zone of late-type M dwarfs \citep{gao15}.

On planets with a significant water reservoir, abiotic production of O$_2$ and O$_3$ can occur through photolysis
of water vapor with subsequent escape of hydrogen.
\citet{luger15} studied water loss and related oxygen buildup around M-dwarfs.
They used a simple model of water photolysis and hydrogen escape that is energy-limited
by the XUV flux or diffusion-limited when hydrogen had to diffuse through an oxygen atmosphere
that had built up.
They found that terrestrial planets with sufficient surface water could have lost several Earth oceans
in the first few 100 million years after formation due to the extended pre-main sequence phase
of their host star.
Tracking the evolution of the water content of the planet and the oxygen content of its atmosphere
through geologic times, they showed that these processes could build up abiotic oxygen atmospheres
of hundreds to thousands of bars.

Water loss on terrestrial exoplanets can occur around any host star if water vapor
can reach the upper atmosphere such that it is available for photolysis and subsequent hydrogen escape.
Whether this process is effective depends on the atmospheric temperature structure, which
is controlled by the stellar insolation as well as the atmospheric composition \citep{kasting88,wordsworth13}.
{Atmospheres of sufficient thickness develop a tropopause or 'cold-trap' that separates the lower
atmosphere, which is controlled by convective processes and in which the temperature gradient
is close to an adiabat, from the middle atmosphere, which is controlled by the absorption of short-wave solar radiation,
leading to temperatures being close to isothermal or even increasing with altitude.
The tropopause typically occurs at pressures of order 0.1 bar as the atmosphere becomes opaque to thermal radiation
at higher pressures, causing temperatures to increase at lower altitudes.
This dependence has been shown to occur on Earth, Venus, Titan, and the giant planets,
and is also expected to occur in exoplanetary atmospheres \citep{robinson_catling14}.}
The influence of the atmospheric composition on tropopause formation is not limited to radiative gases like CO$_2$.
\citet{wordsworth14} showed that in a moist convective environment
the temperature structure and hence the amount of water vapor that reaches the upper atmosphere 
depends largely on the amount of non-condensable gases in the atmosphere.
The amount of non-condensables determines whether an atmosphere can develop a 'cold-trap'
that contains most of the water in the lower atmosphere
and dries out the upper atmosphere.
If an atmosphere is low in non-condensables, water vapor becomes a major constituent and
the cold-trapping is inhibited, leading to a much moister upper atmosphere.
For terrestrial exoplanets the primary non-condensable is molecular nitrogen but it also could be a noble gas like
argon. \citet{wordsworth14} showed that atmospheres with certain combinations of N$_2$ and CO$_2$ amounts
can lead to surface temperatures close to Earth's average surface temperature of 288 K
but have upper atmospheric water vapor mixing ratios that differ by several orders of magnitude.

Water vapor in the upper atmosphere is photolyzed due to the availability of hard UV radiation
yielding hydrogen and oxygen. The light hydrogen will escape to space while the oxygen will stay behind.
This will lead to buildup of oxygen in the atmosphere, which is buffered by the oxidation
of the surface or the interior of the planet. For exoplanets these oxidation rates could vary
over orders of magnitude depending on the initial oxidation state of the surface or the interior,
and on whether the surface is connected to the interior through a process like plate tectonics.
This would probably be true even for ocean planets without land masses unless the ocean is deep
enough to form high-pressure ices on the sea floor that prevent an exchange of oxygen with the
surface \citep{kaltenegger13}.
{\citet{wordsworth18} recently showed that in an early evolutionary phase of a planet dominated
by a magma ocean, abiotic oxygen can be absorbed very rapidly by the mantle due to the reaction with
FeO. However, the volatile inventory of a planet is affected by a multitude of factors over its geologic history,
including possible delivery after planet formation as has been debated for Earth \citep{hartogh11}.}

The buildup of O$_2$ ends when all water is lost or when
oxygen has built up to a level when it is a major non-condensable such that it facilitates a
cold-trap for water vapor. Even transitional states in this process might exist long enough
to have a certain probability of being observed.

Here we use a one-dimensional coupled photochemistry/radiative-convective equilibrium model to
model the temperature and water vapor structure as well as the chemical composition of exoplanetary
atmospheres with N$_2$, CO$_2$, and H$_2$O as major atmospheric constituents.
We investigate the influence of the ratio of these constituents on structure and composition.
Our approach allows us to interactively model the atmospheric structure
in response to changes in atmospheric composition,
and in turn the atmospheric composition
in response to changes in atmospheric structure in a moist convective regime.
We quantify the buildup of abiotic O$_2$ over geologic time scales based on water vapor photolysis
and subsequent hydrogen escape.
We calculate the amount of ozone that is photochemically produced in such atmospheres
and quantify its effect on the temperature structure.
Finally, we use a radiative transfer model to calculate near-IR and mid-IR spectra for transits of such exoplanets
and identify key spectral features with a focus on the capabilities of the James Webb Space Telescope (JWST).

\section{Model Descriptions}
\label{models}

Our approach enables a systematic survey of the parameter space 
for spectral signatures of oxidized exoplanetary atmospheres. It relies on 
the combination of three well developed and tested models: a chemical kinetics model 
of planetary atmospheres which provides the abundance profiles of the atmospheric constituents, 
a radiative-convective equilibrium code to self-consistently calculate 
the atmospheric temperature and pressure structure and a radiative transfer code 
to create synthetic spectra to show how these atmospheres will look when observed.

\subsection{Photochemical model}
\label{kinetics}

We model the chemistry of the exoplanet atmosphere using the
Caltech/JPL coupled photochemistry/dynamics code KINETICS \citep{allen81,yung_demore99}.
KINETICS is a fully implicit finite difference code that solves the coupled continuity equations 
for each species and includes transport via molecular and eddy diffusion. It has been used previously 
for modeling exoplanet atmospheres \citep{line11,moses11,moses13,gao15}. For our purposes, 
the code is used in its 1D mode, allowing the distribution of molecules to be determined in a 
vertical column from the surface of the planet to the top of the atmosphere.

We calculate the chemistry
of 31 species linked by 175 reactions focusing on the nitrogen, oxygen
and hydrogen species on an altitude grid with 1-2 km spacing.
The reaction network includes the condensation
and evaporation of water onto existing aerosol particles. \citet{willacy16}
provides a detailed description of the treatment of these
processes, but in brief the condensation rate is calculated from the
rate of collision of gas species with the aerosol particles, and the
evaporation rate depends on the saturation vapor pressure. The
treatment is numerically stable.

The diurnally averaged radiation field is calculated at a latitude of
$30^\circ$, assuming a planetary day with an arbitrary length of 24 hours.
The planet is assumed to be located at 1 AU and the flux is solar.

\begin{figure}
\plotone{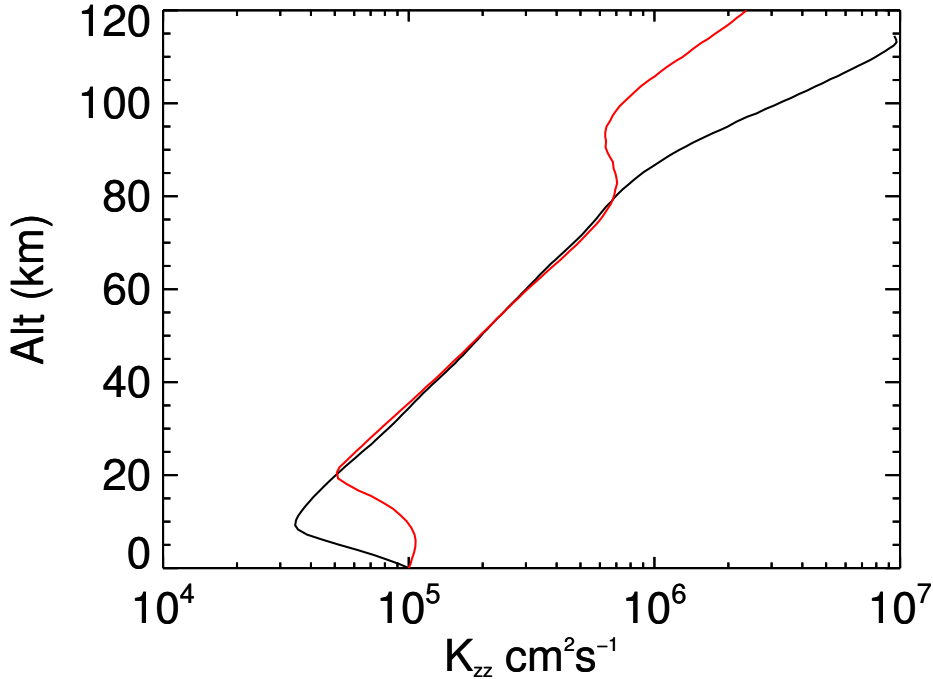}
\caption{Profiles of K$_{zz}$ for an atmosphere with Earth's N$_2$ inventory
(black) and for an atmosphere with 20\% of Earth's N$_2$ inventory (red).
The dry molar CO$_2$ mixing ratio is 400 ppm in each case.}
\label{kzz}
\end{figure}

KINETICS includes transport by both molecular and eddy diffusion. The
value of the vertical diffusion coefficient is uncertain in exoplanetary atmospheres.
Here we calculate the eddy diffusion coefficient as a function of altitude
assuming free convection \citep{gierasch_conrath85}.

\begin{equation}
K_{zz} = \frac{H}{3} \left( \frac{L}{H} \right)^{4/3} \left( \frac{R \sigma T^4}{\mu \rho c_p} \right)^{1/3},
\end{equation}
where H is the scale height given by $\frac{R T}{\mu g}$, R is the
universal gas constant, $\sigma$ is the Stefan-Boltzmann constant,
$\mu$ is the atmospheric molecular weight, $\rho$ is the atmospheric
mass density, $c_p$ is the atmospheric heat capacity. L is the
mixing length defined with respect to the convective stability of the
atmosphere \citep{ackerman_marley01} as

\begin{equation}
L = H max(0.1, \Gamma/\Gamma_a),
\end{equation}
where $\Gamma$ is the lapse rate and $\Gamma_a$ is the adiabatic lapse
rate.  Following  \citet{ackerman_marley01} a lower bound for L is set
at 0.1 H in the radiative regions of the atmosphere.
Following \citet{gao15} we scale the diffusion so that the value at
the planet surface is $10^5 \mathrm{cm}^{-2} \mathrm{s}^{-1}$. The
excess K$_{zz}$ is likely a result of the the equations' assumption of free
convection breaking down in the presence of a solid surface.    
Figure \ref{kzz} shows profiles K$_{zz}$ for an atmosphere with Earth's
N$_2$ inventory and for an atmosphere with 20\% of Earth's N$_2$ inventory.
Differences are most noticeable in the lower atmosphere due to different
lapse rates, as well as in the upper atmosphere. In the middle atmosphere
the K$_{zz}$ values are very similar as they are dominated by
the lower bound of the mixing length.

Photolysis of water vapor will lead to the formation of hydrogen and
oxygen in the upper atmosphere. In the model H and H$_2$ can escape from the
atmosphere. We estimate the hydrogen escape rate assuming diffusion-limited escape.
{This concept has been successfully applied to Earth, Mars, Venus, and Titan (e.g. \citet{catling_kasting17})
as well as rocky exoplanets \citep{gao15}. The scenarios described in this study are
comparable to Earth as we assume an Earth-size planet
with a nitrogen-dominated atmosphere at comparable surface pressures
around a Sun-like star at 1 AU distance.
Measurements of exospheric temperatures and hydrogen densities show that the total hydrogen escape flux at Earth
is close to the diffusion-limited escape flux \citep{bertaux75}.
Other species are assumed to not escape from the atmosphere,
such that non-thermal loss mechanisms, e.g. for atomic nitrogen and oxygen, are neglected.}

We calculate diffusion-limited hydrogen escape after \citet{hunten73},
where the escape flux $\phi_i$ is given as
\begin{equation}
\phi_i \approx \frac{b_i f_i}{H}
\end{equation}
and the corresponding velocity of the escaping gas is
\begin{equation}
v_i \approx \frac{b_i}{n H} = \frac{D_i}{H}.
\end{equation}
Here D$_i$ is the diffusion coefficient, b$_i$ is the binary collision coefficient,
n is the density of the atmosphere and f$_i$ is the mixing ratio of the escaping species.

At the lower boundary, water is assumed to always be saturated at the
planet's surface. The fate of O$_2$ molecules at the planet's surface
depends on whether they can be removed from the atmosphere,
e.g. potentially by being incorporated into carbonates, FeO or other refractory molecules.
We investigate scenarios with no O$_2$ sequestration at the surface
as well as scenarios within an order of magnitude of the average oxidation rate on Earth
due to Fe$^{3+}$ subduction into the mantle over the last 4 Ga,
estimated at $(1.9-7.1) \cdot 10^9$ molec. O$_2$ $\mathrm{cm}^{-2}\mathrm{s}^{-1}$
\citep{catling01}.

KINETICS has been coupled with the radiative-convective equilibrium code RC1D (Section \ref{rc1d})
to enable changes in the thermal and pressure structure of the atmosphere induced 
by the chemistry.
When KINETICS calls RC1D it passes concentrations, pressures and temperatures at each altitude level to
RC1D. RC1D passes the updated temperatures and concentrations for the new conditions
back to KINETICS. KINETICS then recalculates the altitude grid and the new boundary condition
for water vapor to enforce that it is saturated at the bottom of the atmosphere.
The values of K$_{zz}$ and the escape velocities are recalculated at each
time step as the atmospheric temperature evolves.
{KINETICS uses a variable time step depending on the model time and on how rapidly the chemical changes occur.}
Initially KINETICS calls RC1D every 10 time steps.
Once the model has started to evolve and reached a model time of 5000 model years,
RC1D is called every time step.

\subsection{Radiative-convective equilibrium model}
\label{rc1d}

In order to allow the inclusion of changes in the thermal and pressure structure of the atmosphere induced 
by the chemistry KINETICS has been coupled with a radiative-convective equilibrium code (RC1D).
RC1D is a one-dimensional non-gray radiative-convective equilibrium model.
It uses a modified Uns{\"o}ld-Lucy temperature-correction procedure \citep{lucy64,mihalas78}
to iteratively converge on the radiative equilibrium temperature profile
that produces zero net divergence of the total (stellar- and thermal-wavelength)
radiative flux in the atmosphere at each altitude. Short-wavelength
($\lambda < 5 \mu$m) radiative fluxes are calculated using a two-stream
delta-Eddington approximation \citep{briegleb92}. Thermal radiative fluxes are calculated
using a two-stream Feautrier approach that ensures recovery of the diffusion limit at
large optical depths. Broadband molecular transmittances are calculated for CO$_2$, CH$_4$, H$_2$O, and O$_3$
using the correlated-k approach \citep{lacis91,liou02}.
{Opacities are computed from the
HITRAN 2012 \citep{hitran12} and HITEMP 2010 \citep{hitemp10} line lists
for each selected molecule on a temperature-pressure-wavenumber grid.
We use a wavenumber grid with 10 cm$^{-1}$ spacing at long wavelengths
and 100 cm$^{-1}$ spacing at short wavelengths.
The pressure grid has 22 points from 10$^{-6}$ to 10 bar on a logarithmic scale,
and the temperature grid has 19 points from 87 to 303 K with 12 K spacing.
The algorithm sorts the opacities within each wavenumber interval at each
specific temperature and pressure to generate opacity vs. cumulative probability functions.
Finally, the routine computes a set of 20 Gaussian-quadrature opacity values that compose the integrated opacity for the wavenumber interval.}
The treatment of opacity also includes Rayleigh scattering as well as aerosol scattering,
if required \citep{friedson09}. The temperature-dependent opacities are updated during
each iteration of the temperature correction procedure.

A convective adjustment is applied whenever the radiative-equilibrium lapse rate
exceeds the local moist adiabat. Calculation of the moist adiabatic lapse rate
is based on the method discussed by \citet{kasting88}, which is designed to provide
the correct adiabat irrespective of the water inventory in the atmosphere.
Analytical expressions for the temperature dependent heat capacities of
N$_2$, CO$_2$ and H$_2$O were included analogously to \citet{wordsworth13}.
To account for molecular oxygen as a major atmospheric constituent,
the following expression for the heat capacity of O$_2$ based on \citet{chase98} was added
\begin{equation}
c_{p,O_2} = 1118.1 - 0.722 T + 2.07\cdot10^{-3} T^2 - 1.30\cdot10^{-6} T^3 - 2.63\cdot10^5 T^{-2} J kg^{-1} K^{-1}.
\end{equation}
Since performing a moist adjustment disturbs the calculated radiative equilibrium
at levels outside the adjusted region, the process of calculating
a radiative-convective equilibrium profile is done iteratively.
The process continues until
\begin{enumerate}
\item all unstable layers have been adjusted to a moist adiabatic lapse rate;
\item all stably stratified levels are in radiative equilibrium
(to within a tolerance of typically less than 0.1\% disagreement
between the net solar and net infrared fluxes at any level);
\item the net outgoing longwave radiation equals the net stellar energy entering the atmosphere. 
\end{enumerate}

\subsection{Radiative transfer model}
\label{ods}

Simulated transit spectra are calculated with the radiative transfer code ODS
(Occultation Data System) \citep{mccleese92}. ODS is a state-of-the-art line-by-line program 
that calculates absorption coefficients, transmission, and emission spectra for limb and on-planet 
viewing geometries. Calculations are performed for all lines with no approximations. 
This has the advantage over other methods (e.g. non-overlapping line approaches, correlated-k) 
that there are no implicit assumptions of correlations between absorptions that may be specific 
to a certain atmospheric composition, or only valid in certain pressure or temperature ranges. 
This makes the code very versatile so that spectra from a wide range of exoplanets can be calculated. 
A line-strength cutoff can be utilized to speed up calculations by omitting weak lines once 
it has been determined that they do not contribute significantly to the result. 
The code works currently with the HITRAN 2012 database \citep{hitran12}. The code uses a look-up table for 
the Voigt lineshape and accounts for the sub-Lorentzian lineshape of CO$_2$-lines \citep{burch69}. 
The CKD water vapor continuum \citep{ckd89} is included into the code
to account for water vapor continuum contributions.
ODS has been used to calculated atmospheric transmissions and emissions for studies 
of Venus, Earth, and Mars. It is currently used to calculate transmission tables 
for retrievals of atmospheric parameters from Mars Climate Sounder 
limb and nadir observations at Mars \citep{kleinboehl09}. The code has been validated 
against the NEMESIS radiative transfer code \citep{irwin08}.

The ODS code is readily capable of calculating limb transmissions of the atmosphere at different tangent altitudes. 
It integrates these transmissions over altitude in order to calculate the total effective radius,
and in turn the total effective area, of the exoplanet at each wavelength. The ratio of the
total effective area of the planet to the disc area of the host star corresponds to the signal
that is measured by an observer during a primary transit \citep{irwin14}.

\section{Results and Discussion}
\label{results}

Atmospheric profiles of temperature and chemical constituents were calculated
for a wide range of input parameters using the combined KINETICS/RC1D models.
In each case the planetary radius was set to Earth's radius and the insolation
was assumed to be from a Sun-like star in 1 AU distance. Simulations were performed
for atmospheric N$_2$ inventories at 20\%, 40\%, 60\%, 80\% and 100\% of Earth's
atmospheric N$_2$ inventory. For each N$_2$ inventory several simulations were performed
with CO$_2$ dry molar mixing ratios between 100 ppm and 1\%. Model simulations were run 
for up to 1 billion model years.
Initial simulations do not consider an oxygen sink through oxidation of the surface, 
and hence allow oxygen produced by water vapor photolysis with subsequent hydrogen escape 
to build up in the atmosphere.

\begin{figure}
\plotone{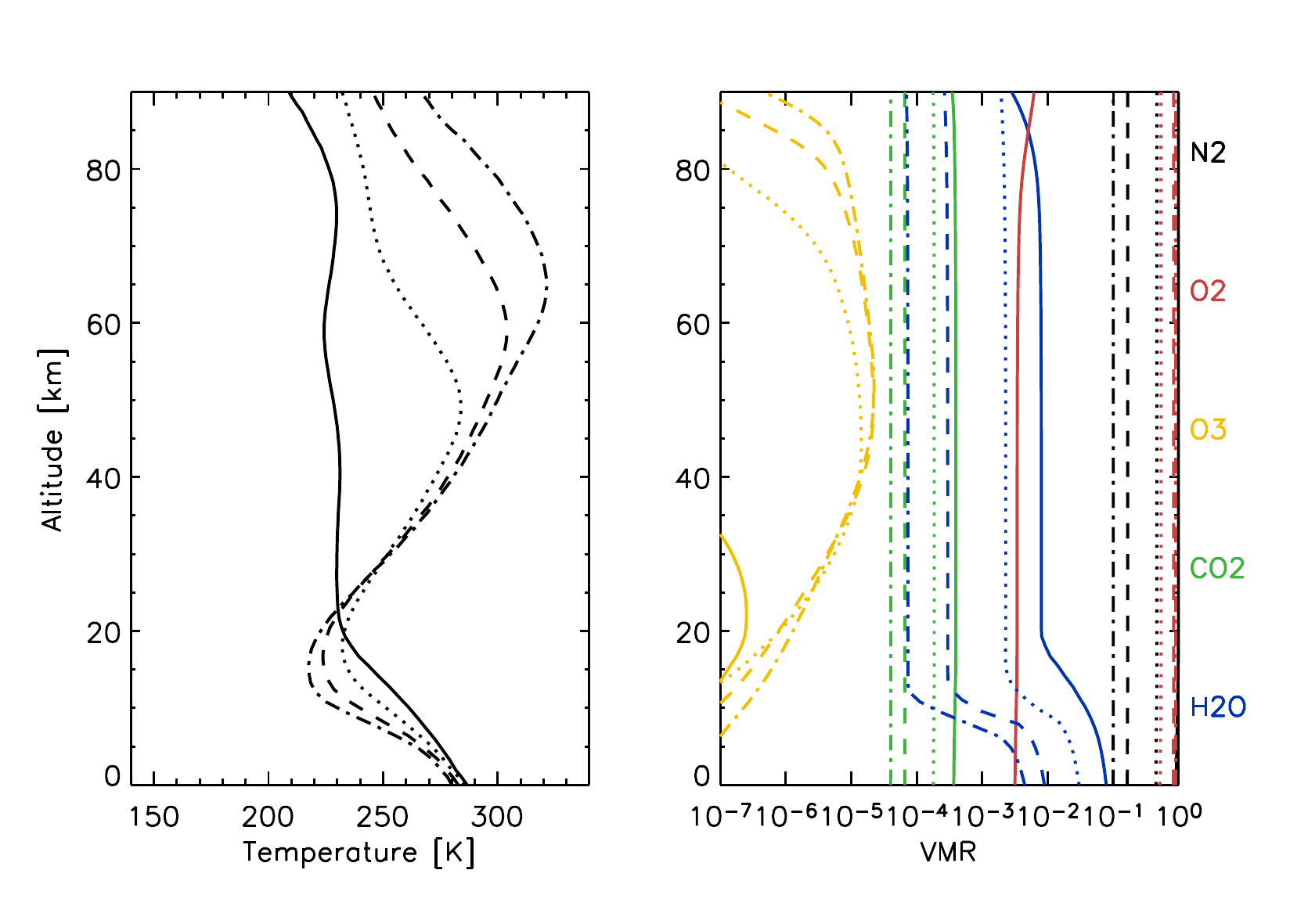}
\caption{Atmospheric profiles of temperature (left) and chemical constituents
(right) with 20\% of Earth's N$_2$ inventory, a dry molar CO$_2$ mixing ratio of 400 ppm,
and no surface sink for O$_2$. Model output is given for 10000 (solid), 
10 million (dotted), 100 million (dashed) and 1 billion model years (dash-dotted).}
\label{n202}
\end{figure}

The key influence of the amount of non-condensables on the atmospheric lapse rate
is shown in Figure \ref{n202}.
It shows the output of a model run with 20\% of Earth's N$_2$ inventory
and a dry molar CO$_2$ mixing ratio of 400 ppm. The temperature profile
in the lower atmosphere is dominated by moist convection and follows 
a moist adiabatic lapse rate until it reaches a cold-trap at 22 km altitude.
In analogy to Earth we call this layer the troposphere and the cold-trap region the tropopause.
The tropopause temperature controls the amount of water vapor that is able to enter 
the region above the tropopause, which we call in analogy to Earth the stratosphere.
H and H$_2$ obtained from water vapor photolysis in the middle atmosphere is lost 
through diffusion-limited escape as given in Section \ref{kinetics}.
In the beginning of the simulation, after 10000 model years, the lapse rate is very shallow 
due to the low N$_2$ inventory. While the surface temperature is
287 K the tropopause temperature is only 231 K at a tropopause altitude of 22 km, 
corresponding to a lapse rate of only about 2.5 K/km. 
Controlled by the tropopause temperature, water vapor enters the stratosphere 
with a mixing ratio of order 1\%. It is constant throughout the stratosphere 
and only starts to decrease at high altitudes
due to photolysis. At this early stage in the model run, O$_2$ has not had a chance to built up to
to significant amounts and is nearly constant at 0.3\%, while its photochemical product O$_3$
has a maximum mixing ratio of about 0.3 ppm around 20 km altitude. 
{We note that these middle atmospheric temperatures tend to be
somewhat higher than in cases with intermediate to high N$_2$ inventories by \citet{wordsworth14},
which is likely related to differences in the model top and the details of the radiative transfer.
The middle atmospheric temperature minima in Figure \ref{n202} are close to the skin temperature
of an Earth-like planet at 1 AU, suggesting that the radiative characteristics are not too far from a gray absorber.
In addition, the production of ozone, which is already noticeable after 10000 model years,
has a radiative effect that impacts temperatures around the tropopause and dominates
middle atmospheric temperatures later in the evolution.}

Significant atmospheric changes are seen after 10 million model years 
(dotted lines in Figure \ref{n202}). Oxygen now has built up to a level comparable 
to the amount of nitrogen in the atmosphere and similarly to the nitrogen, 
it acts as a non-condensable in the atmosphere. We note that the partial pressure of oxygen 
at this stage of the simulation is comparable to oxygen on Earth. 
In response, the tropospheric lapse rate has steepened, such that 
the tropopause altitude  has dropped to 17 km. Water vapor now enters the stratosphere
at a mixing ratio of about 0.3\%. 
Oxygen photochemistry leads to the production of O$_3$ at levels around 10 ppm, 
which is comparable to the mixing ratio levels typically found in Earth's atmosphere.
These levels of ozone cause significant heating of the stratosphere due to absorption of UV light,
leading to maximum stratospheric temperatures of nearly 290 K around 50 km.

\begin{figure}
\plotone{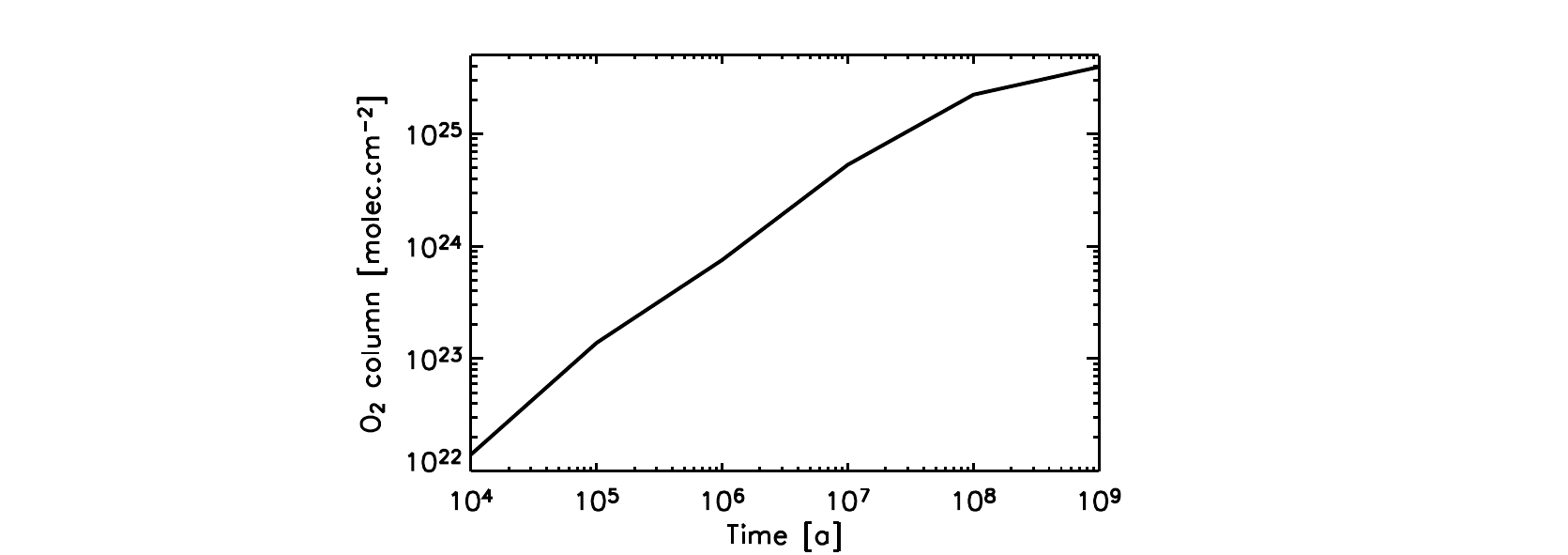}
\caption{Increase in column O$_2$ over geologic timescales based on the simulations in Figure \ref{n202}.}
\label{o2col}
\end{figure}

At 100 million model years (dashed lines in Figure \ref{n202}) oxygen has 
continued to build up and now significantly exceeds the amount of nitrogen 
in the atmosphere. The increase in non-condensables leads to a further steepening 
of the atmospheric lapse rate, such that the tropopause temperature is now 224 K 
and the stratospheric water vapor mixing ratio has dropped to $\sim$400 ppm. 
Oxygen photochemistry leads to ozone levels of order 25 ppm.

After 1 billion model years (dash-dotted line in Figure \ref{n202}), oxygen has 
build up to a mixing ratio of nearly 90\% in the atmosphere, leading to a surface pressure 
of about 1.7 bar and a nitrogen mixing ratio of about 10\%. The tropopause is now located 
at 14 km with a temperature of $\sim$218 K. With a surface temperature of 280 K 
this corresponds to a lapse rate of 4.4 K/km. For comparison, the average lapse rate in
Earth's troposphere is 6.5 K/km between 0 and 11 km altitude \citep{us76}.
The stratospheric water vapor levels are of order 70 ppm. 
Note that over the billion year time span considered in the model, 
a ninefold increase in surface pressure due to the buildup of abiotic oxygen 
has led to the drying of the stratosphere by over two orders of magnitude. 
Ozone levels in the middle atmosphere are now of order 30 ppm,
leading to temperatures of $\sim$320 K around 65 km in the middle atmosphere.

{Figure \ref{o2col} shows the increase in the O$_2$ column over 1 billion years
for the model scenario given in Figure \ref{n202}. While the increase is nearly linear
between 100000 and 10 million model years, it significantly slows down
after 10 million years and even more after 100 million years as the water vapor supply
to the upper atmosphere is reduced due to the influence of the oxygen on the lapse rate.
After 1 billion model years molecular oxygen has built up to a column amount of about
$4 \cdot 10^{25} \mathrm{molec.cm}^{-2}$.
}

\begin{figure}
\plotone{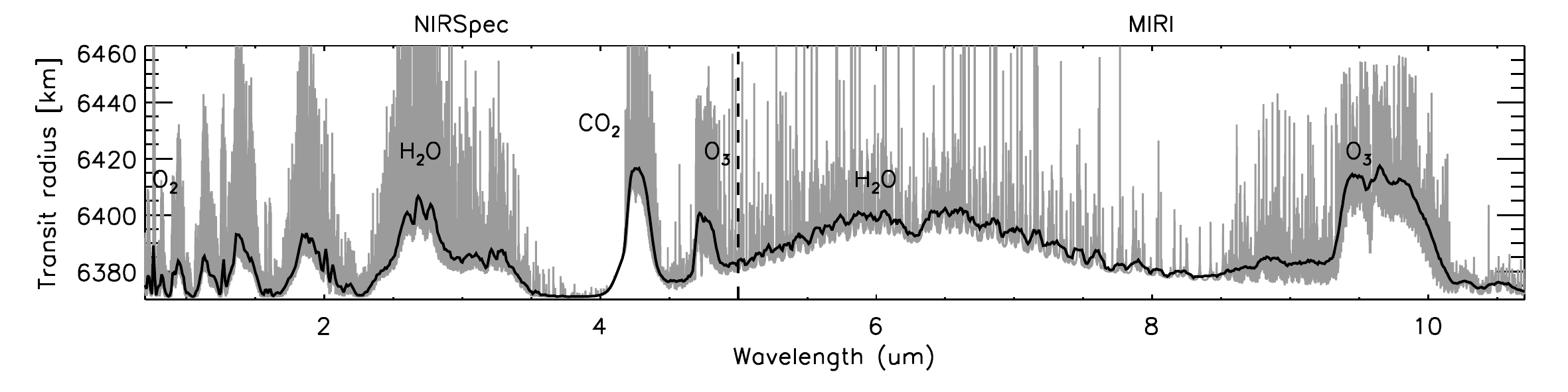}
\plotone{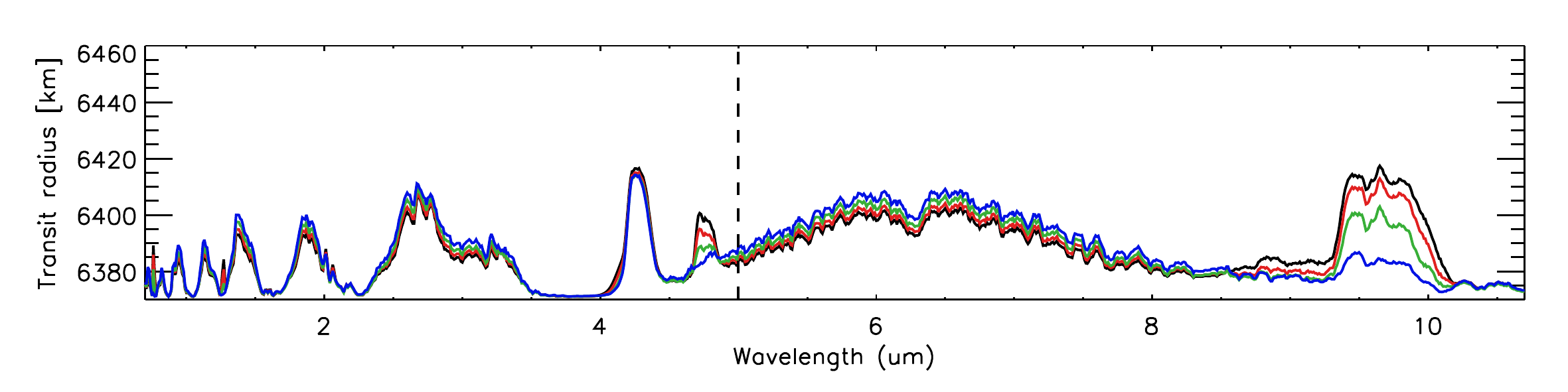}
\plotone{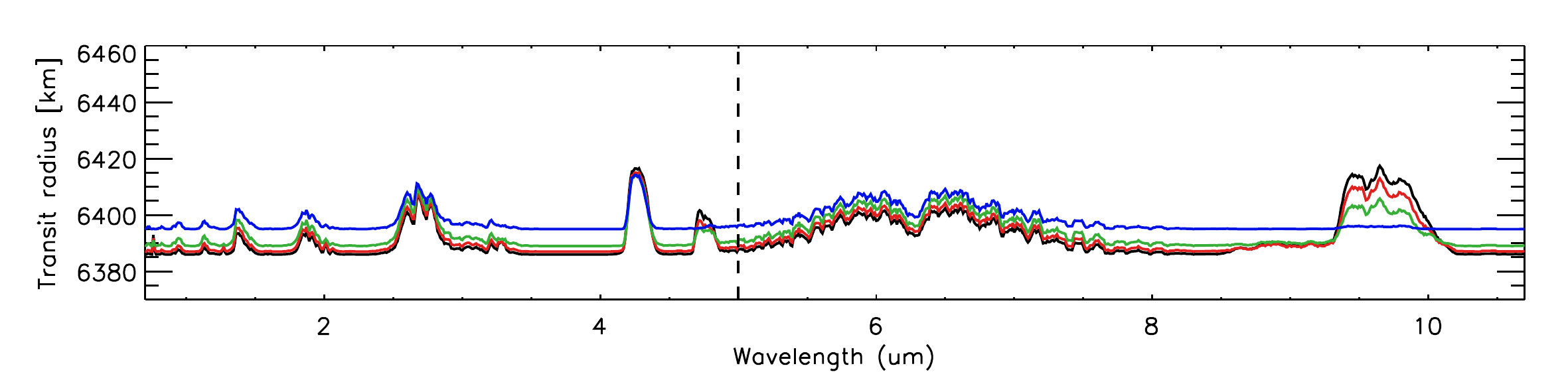}
\caption{Spectrally resolved apparent radius of a transiting exoplanet 
with the atmospheric structures and compositions based on Figure \ref{n202}. 
The top panel shows the case after 1 billion model years 
with the gray lines giving the extinction spectrum at high spectral resolution, 
while the black line shows the spectrum convolved to a resolving power of 100, 
comparable to the low resolution modes of the NIRSpec and MIRI instruments on JWST. 
Major molecular absorption bands are identified. 
The middle panel shows a comparison of low resolution transit spectra 
based at the atmospheric conditions after 10000 (blue), 10 million (green), 
100 million (red) and 1 billion model years (black). 
The bottom panel shows the transit spectra for the same atmospheric conditions 
assuming extended cloud cover in the troposphere.}
\label{n202spec}
\end{figure}

We investigate how these atmospheres would look in a transit observation. Figure \ref{n202spec}
shows results results of radiative transfer calculations with the ODS code and the profiles from 
Figure \ref{n202} as input. Calculations were performed at high resolution (gray lines in the top panel
of Figure \ref{n202spec}) over a range from 0.6 to 11 $\mu$m. The range up to 5 $\mu$m is covered by
the Near-Infrared Spectrograph (NIRSpec, \citet{rauscher07}) while the range beyond 5 $\mu$m is covered by
the Mid-Infrared Instrument (MIRI, \citet{wright04}) on JWST.
The high resolution spectrum is convolved with a running mean to simulate a spectrum with a resolving
power of order 100, which corresponds to the low resolution modes of the NIRSpec and MIRI instruments.
We note that the wavelength range between 0.6 to 2.8 $\mu$m is also covered by the
Near-Infrared Imager and Slitless Spectrograph (NIRISS, \citet{doyen12}) on JWST,
which can achieve resolving powers between 150 and 700 with slitless spectroscopy.

The spectrum corresponding to an early atmosphere (10000 model years, blue line in middle panel
of Figure \ref{n202spec}) is dominated by absorption bands of water vapor and CO$_2$.
The CO$_2$ amount in the atmosphere leads to a transit height of 43 km at 4.3 $\mu$m
at this spectral resolution. The water vapor amount of $\sim$1\% in the middle atmosphere leads
to transit heights of 38-40 km in both the spectral regions around 2.7 and 6.5 $\mu$m. Even at levels
of less than a ppm an ozone absorption feature is already discernible at 9.6 $\mu$m.
The transit spectra change as the atmospheric simulation progresses over geologic time scales.
While the CO$_2$ feature stays virtually constant over time, the decrease in middle atmospheric
water vapor leads to a significant reduction in the transit radii caused by the water features.
After 1 billion model years the water feature at 6.5 $\mu$m has decreased to a height of 31 km.
The water feature at 2.7 $\mu$m has maximum heights of 29 km. The buildup of atmospheric
ozone over the time of this simulation leads to a strong ozone absorption at 9.6 $\mu$m.
Even though the maximum ozone mixing ratio is of similar magnitude in the simulations stages
at 10 million, 100 million and 1 billion model years
the different vertical extend leads to transit heights of 32, 42 and 46 km,
respectively. Notable is also the rise of the oxygen A-band at 0.76 $\mu$m as the oxygen builds up
in the atmosphere throughout the simulation. At a resolving power of 100 the apparent radius rises
from 10 km at 10 million years to 15 km at 100 million years, and to 18 km at 1 billion years.

An interesting aspect to consider is the effect of tropospheric clouds on the transit spectrum.
If the troposphere of an exoplanet is moist convective, it will likely allow the formation of
widespread cloud cover. While the radiative feedback of potential cloud cover on the thermal structure
of the atmosphere is complicated and has not been included at this stage of our simulations,
it is easy to test the effect of tropospheric clouds on the transit spectrum. Clouds are considered
in the transit simulation as an extinction (0.1 km$^{-1}$) by moderately large (8 $\mu$m effective radius)
water ice particles throughout the troposphere. Due to the limb geometry that is
realized in a transit observation, aerosol particles will either absorb a light ray or scatter it out
of the field of view, such that extinction is the dominating effect on a light ray in a transit
observation. The long atmospheric light path across the planet's limb will cause aerosol layers of
even moderate density to appear opaque in a transit observation. With the spectral signature of
moderately large ice particles being essentially flat, the part of the gas absorption spectrum is largely
determined by the vertical extend of the clouds, which in turn is determined by the tropopause height.
Due to the shallow lapse rate driven by the low amount of non-condensables at the beginning of the
simulation, the tropopause is found around 22 km altitude, which means that the part of the spectrum below
this altitude is essentially unaccessible to the observer. At 10000 model years, the water features
at 2.7 and 6.5 $\mu$m and the CO$_2$ feature at 4.3 $\mu$m will still be accessible. The increase in
the lapse rate caused by the buildup of abiotic oxygen causes the tropopause to move down in altitude.
At 10 million model years it is located around 17 km and at 100 million model years and after,
it is found at around 14-15 km. The transit radii in spectral regions without much gas absorption
(e.g. 3.5-4 $\mu$m) decrease accordingly. At the later stages of the simulation, features of water vapor,
CO$_2$ and ozone will be discernible. Even the oxygen A-band may be identified in conditions
corresponding to 100 million model years and later. However, a problem that may arise when interpreting
real transit observations is that the planetary radius might not be known accurately, and the opaque
cloud cover might be mistaken as the planetary radius. Note that for example the water feature at
6.5 $\mu$m decreased in transit radius by about 9 km between 10000 and 1 billion model years, in the
same time frame the altitude of the opaque cloud layer decreased by 8 km. If the altitude of the cloud
layer were to be interpreted as the planetary radius, the water amount retrieved
for these very different atmosphere might be comparable, and would likely be underestimated.

\begin{figure}
\plottwo{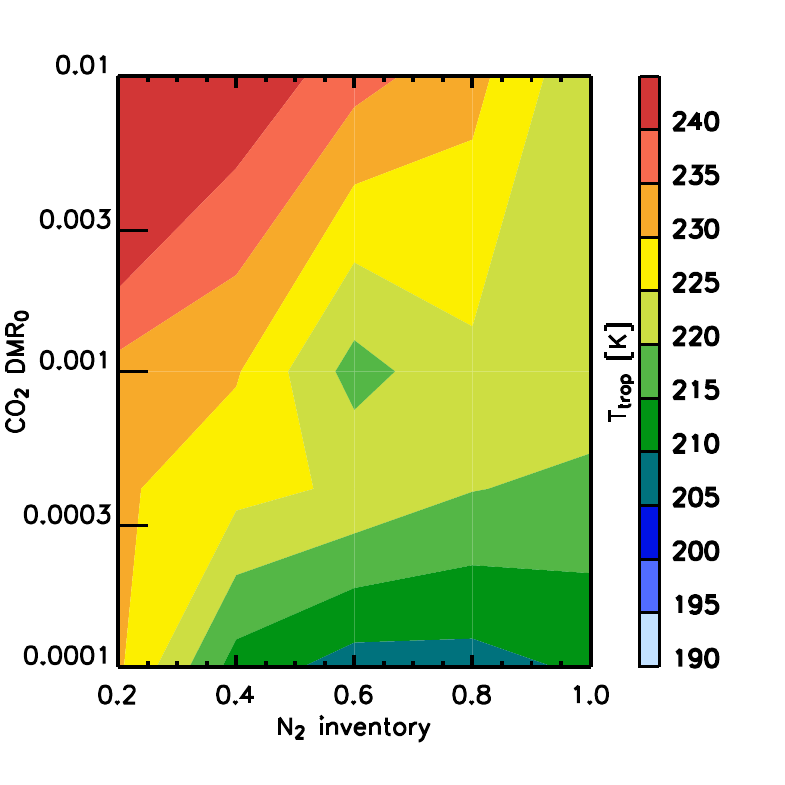}{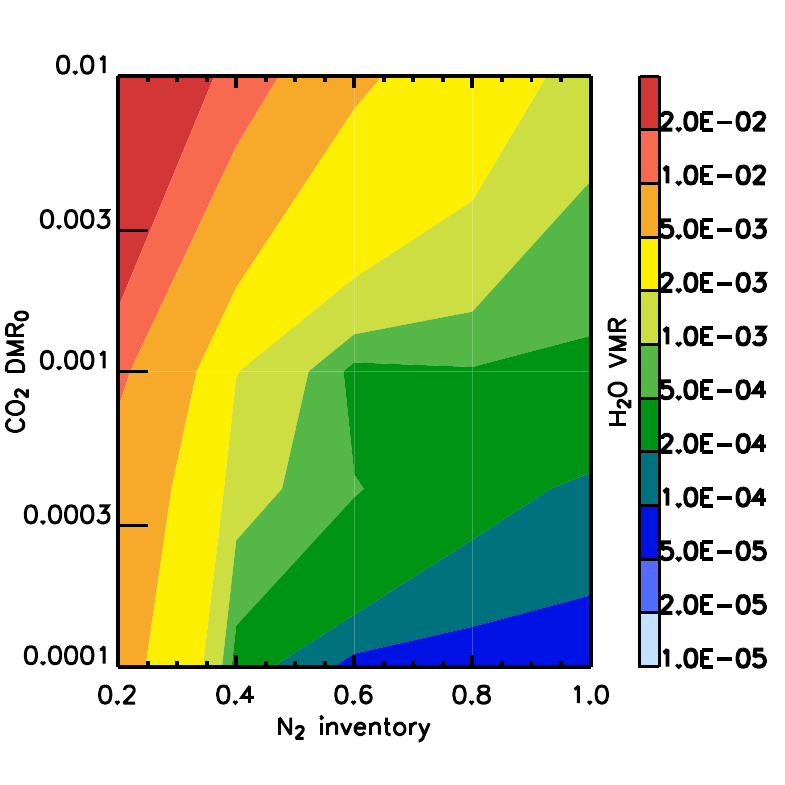}
\caption{Left: Tropopause temperatures for atmospheric scenarios calculated with the combined
photochemistry/radiative convective model for various combinations of N$_2$ inventories and
CO$_2$ initial dry molar ratios after 10000 model years. Right: Modeled stratospheric water vapor
mixing ratios after 10000 model years.}
\label{th2o1e4}
\end{figure}

The general influence of the amounts of nitrogen and carbon dioxide on the tropopause temperature
and the stratospheric water mixing ratio is shown in Figure \ref{th2o1e4}.
On the low end of the studied nitrogen and CO$_2$ ranges we obtain tropopause temperatures around 230 K
as seen in the example in Figure \ref{n202}. Increasing the nitrogen inventory
while keeping the CO$_2$ low depresses the tropopause temperature to $\sim$210 K for a CO$_2$
mixing ratio of 100 ppm.
This transition happens rather quickly between 40\% and 60\% of EarthÕs nitrogen inventory.
The stratospheric water vapor mixing ratio varies accordingly between nearly 1\% for the low nitrogen
and 50-100 ppm for the higher nitrogen inventories.
An increase in CO$_2$ {raises the temperatures and the water vapor mixing ratios in the lower atmosphere.
This in turn leads to a shallower lapse rate and hence} an increase in tropopause temperature for all nitrogen inventories.
{This behavior is qualitatively consistent with results from 3D climate models investigating the effect of
doubling CO$_2$ in Earth's atmosphere \citep{rind98,fomichev07}.}
For dry molar ratios above 0.1\% tropopause temperatures exceed 220 K for high
and 230 K for low nitrogen inventories. Accordingly,
the stratospheric water vapor mixing ratio increases to a few tenths of a percent in the former
and to a few percent in the latter scenarios.

\begin{figure}
\plottwo{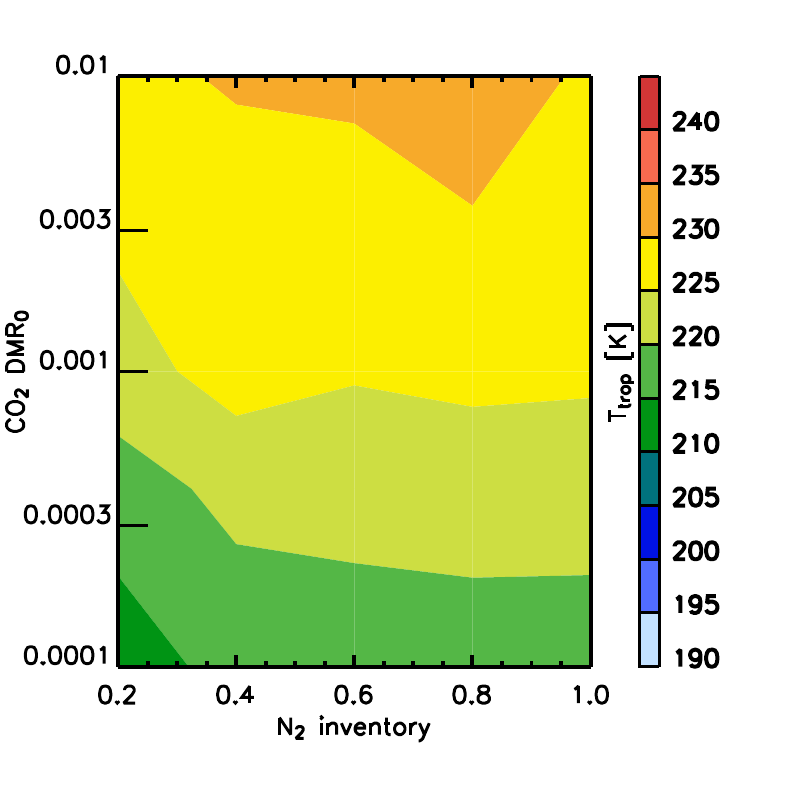}{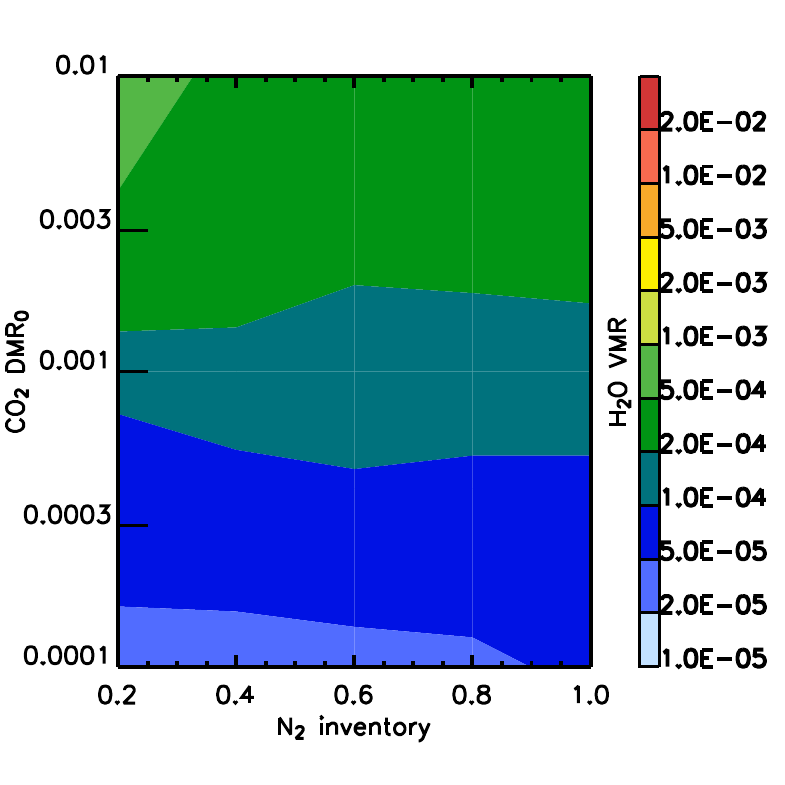}
\plottwo{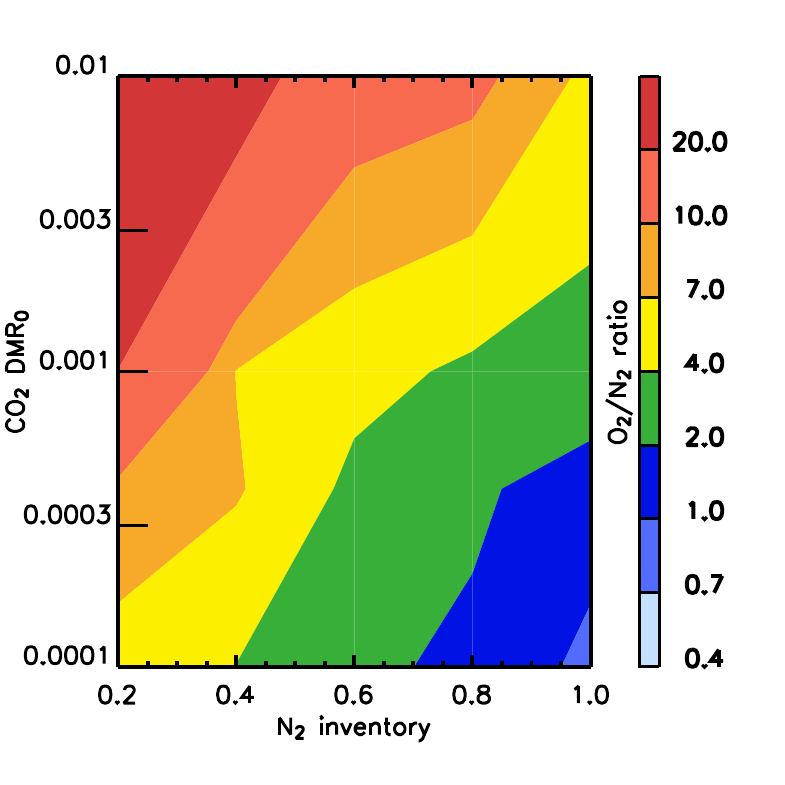}{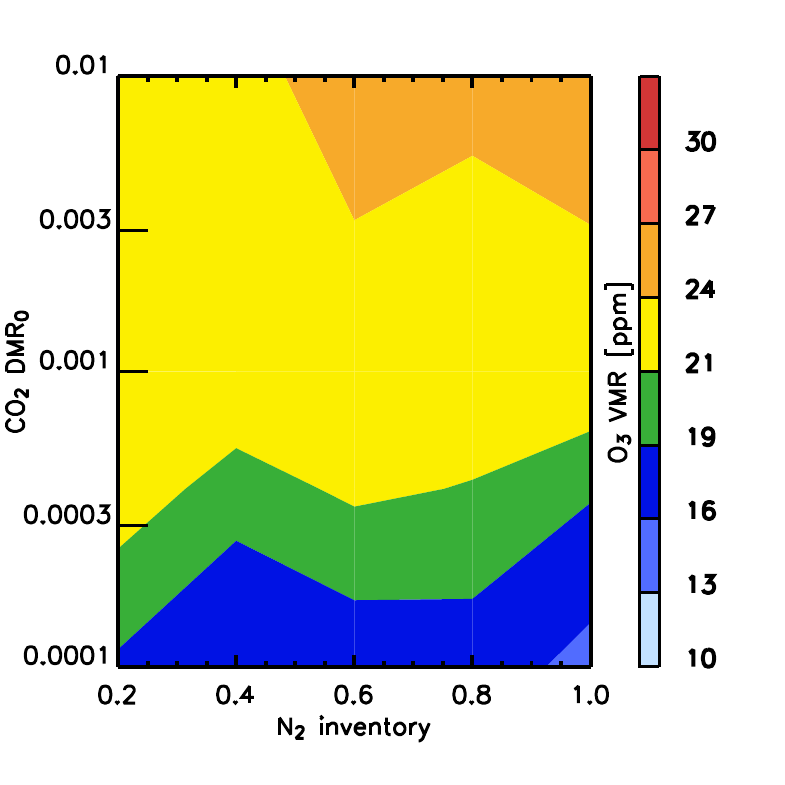}
\caption{Atmospheric parameters modeled with the combined photochemistry/radiative convective model
for various combinations of N$_2$ inventories and CO$_2$ initial dry molar ratios after 1 billion model years.
Top left: Tropopause temperature; top right: Stratospheric water vapor volume mixing ratio;
bottom left: Ratio of O$_2$ to N$_2$ mixing ratios; bottom right: Maximum ozone volume mixing ratio.}
\label{th2o1e9}
\end{figure}

Figure \ref{th2o1e9} shows how these scenarios have developed after 1 billion model years.
For cases with low nitrogen and CO$_2$ inventories, tropopause temperatures dropped significantly
due to the increase in the non-condensable oxygen. In turn, the stratospheric water vapor mixing ratios
have decreased to 100 ppm or less. For scenarios with higher CO$_2$ molar ratio water vapor mixing ratios
also decreased significantly to $\sim$0.1\% or lower.
Figure \ref{th2o1e9} also shows the oxygen and ozone levels that have built up in the atmosphere
after 1 billion model years in absence of a surface sink for oxygen. While oxygen has built up
to significant levels in all scenarios, in the case with the highest nitrogen inventory and
the lowest CO$_2$ levels N$_2$ is still the dominant non-condensable. In cases with lower nitrogen inventories
and/or higher CO$_2$ mixing ratios O$_2$ has exceeded N$_2$ as a non-condensable
and has become the dominant constituent in the atmosphere. In cases with low nitrogen
combined with high CO$_2$, O$_2$ exceeds N$_2$ by a factor of 10 or more. Despite the large
differences in atmospheric oxygen levels the maximum ozone values stay in a range between 10
and 30 ppm. The lowest ozone is found at high N$_2$ inventories and low CO$_2$ levels,
and the highest ozone is found at intermediate to high CO$_2$ levels.
The comparatively dry middle atmosphere after 1 billion years
suggests that further evolution driven by water vapor photolysis and subsequent hydrogen escape
will be rather slow. With durations of hundreds of millions to billions of years, it is likely
that scenarios as described in the aforementioned figures will be realized in exoplanets
without a significant surface sink of oxygen. The duration of these atmospheric states
make it likely that they will be observed at some point. These observations will have to deal
with the ambiguities between potential abiotic and biogenic oxygen production.

\begin{figure}
\plotone{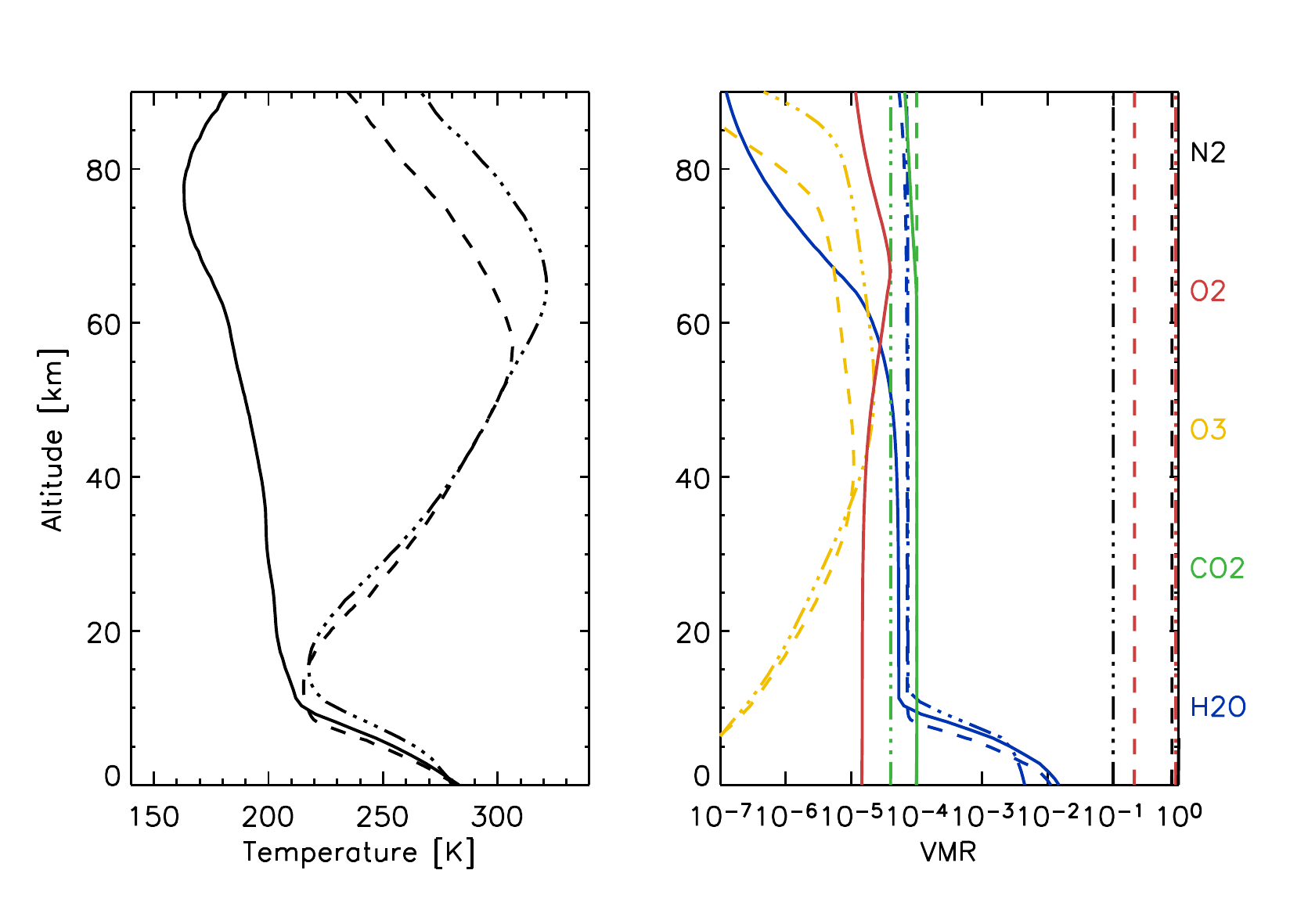}
\caption{Atmospheric profiles of temperature (left) and chemical constituents (right) 
as calculated after 10000 model years with Earth's N$_2$ inventory 
and a dry molar CO$_2$ mixing ratio of 100 ppm (solid lines) 
and with an additional inventory of 21\% O$_2$ assumed to be biogenic (dashed lines). 
This is compared to the scenario with 20\% of Earth's N$_2$ inventory 
and a dry molar CO$_2$ mixing ratio of 400 ppm after 1 billion model years 
(dash-dotted lines, reproduced from Figure \ref{n202}).}
\label{n210o2bio}
\end{figure}

The similarities between the behavior of N$_2$ and O$_2$ as non-condensables in a moist convective
atmosphere are highlighted in Figure \ref{n210o2bio}. It compares temperature and composition profiles
of a run with Earth's N$_2$ inventory after 10000 model years with the results of the run
with 20\% of Earth's N$_2$ inventory after 1 billion model years from Figure \ref{n202}.
n the former case the atmospheric mass is almost exclusively due to nitrogen,
while in the latter case the nitrogen mixing ratio is only about 10\% and the bulk of the atmosphere is oxygen.
The similar heat capacities lead to very similar lapse rates and tropopause temperatures. As a result,
the stratospheric water vapor mixing ratios in both cases are of order 50-100 ppm.
The dashed line in Figure \ref{n210o2bio} shows a run with Earth's N$_2$ inventory
and with an additional inventory of 21\% of oxygen assumed to be of biogenic origin.
This makes this model atmosphere very similar to Earth's atmosphere.
The tropospheric lapse rate and the stratospheric mixing ratio of water vapor are quite comparable
to the run with abiotic oxygen produced over 1 billion years. Ozone is produced with a maximum mixing ratio of
11 ppm, also comparable to the tropical stratosphere of Earth, leading to a significant middle atmospheric
temperature increase.

\begin{figure}
\plotone{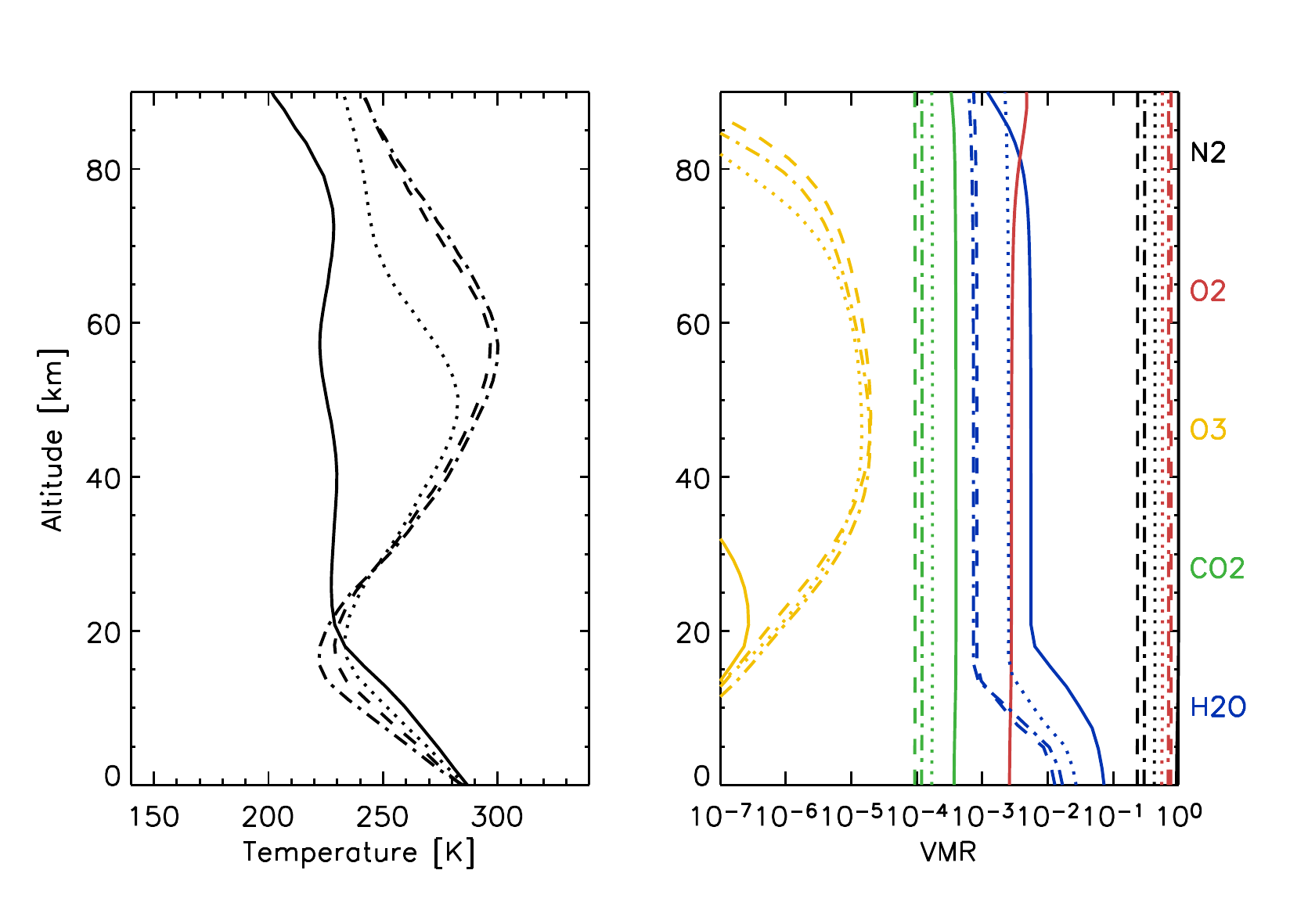}
\caption{Atmospheric profiles of temperature (left) and chemical constituents
(right) with 20\% of Earth's N$_2$ inventory, a dry molar CO$_2$ mixing ratio of 400 ppm,
and an O$_2$ flux to the surface at a rate of
$3 \cdot 10^9$ molec. O$_2$ $\mathrm{cm}^{-2}\mathrm{s}^{-1}$. Model output is given for 10000 (solid), 
10 million (dotted), 100 million (dashed) and 1 billion model years (dash-dotted).}
\label{n202o2flux}
\end{figure}

As it is realistic to assume that some abiotically produced oxygen will contribute to the oxidation
of a planet's surface and mantle, we investigate scenarios with O$_2$ uptake by the surface at rates between
$10^9$ and $10^{10}$ molec. O$_2$ $\mathrm{cm}^{-2}\mathrm{s}^{-1}$.
This encompasses the oxidation rate of Earth due to Fe$^{3+}$ subduction into the mantle,
averaged over the last 4 Ga.

Figure \ref{n202o2flux} shows a run with 20\% of Earth's N$_2$ inventory,
a dry molar CO$_2$ mixing ratio of 400 ppm,
and an O$_2$ flux rate of $3 \cdot 10^9$ molec. O$_2$ $\mathrm{cm}^{-2}\mathrm{s}^{-1}$.
The beginning of the simulation is very similar to the one with the same
N$_2$ and CO$_2$ mixing ratios but without O$_2$ flux to the surface (Figure \ref{n202}).
However, the buildup of abiotic oxygen is slower, and the drying out of the
stratosphere proceeds more slowly as well. After 100 million model years the stratosphere
is dry enough that oxygen production from water photolysis with subsequent hydrogen escape
is largely balanced by the uptake of oxygen to the surface. Changes in temperature and
constituent abundances are only very small between 100 million and 1 billion model years,
indicating that the system is very close to a steady-state. The tropopause is stabilized
around 223 K at an altitude of 18 km, leading to a stratospheric water mixing ratio
of order 700 ppm. While this is significantly higher than in the case without O$_2$ flux
to the surface, it is more than an order of magnitude lower than at the beginning of the
simulation, which started with a stratospheric water mixing ratio of nearly 1\%. The maximum
ozone mixing ratio produced under these conditions is about 20 ppm, which is quite comparable to the
scenario without O$_2$ flux to the surface. Such a planet could exist in this steady state
for geologically extended periods of time. Changes would be expected only due to limits
of the water inventory of the planet or the oxidation capacity of its surface or mantle.
The extended time period over which planets could exist in this state makes such planets
likely candidates of observations.

\begin{figure}
\plottwo{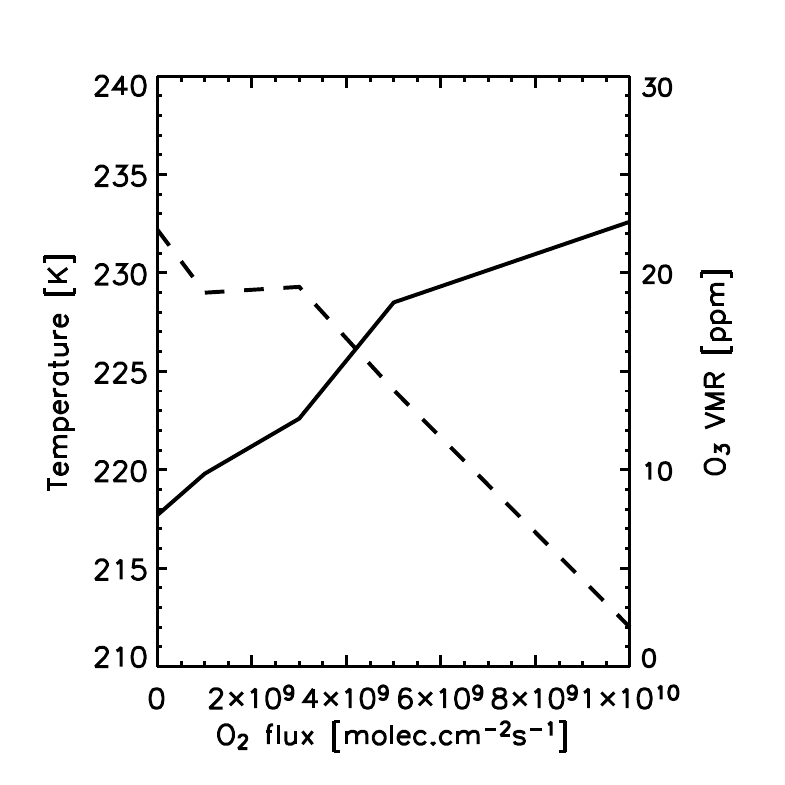}{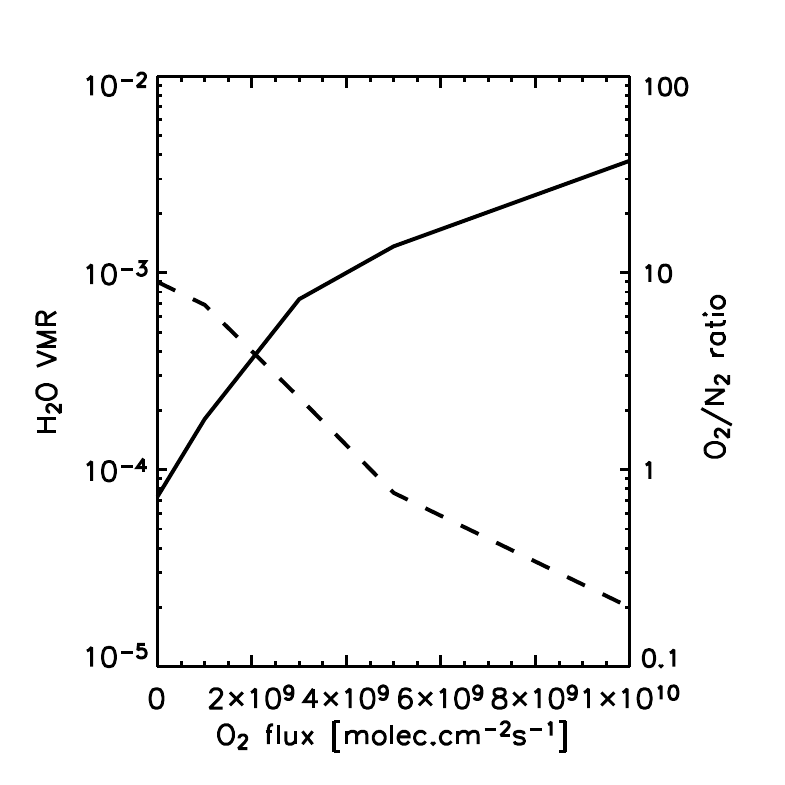}
\caption{Tropopause temperature (left, solid line, left axis),
maximum ozone volume mixing ratio (left, dashed line, right axis),
stratospheric water vapor volume mixing ratio (right, solid line, left axis),
and ratio of O$_2$ to N$_2$ mixing ratios (right, dashed line, right axis)
as modeled with the combined photochemistry/radiative convective model
for 20\% of Earth's N$_2$ inventory, a dry molar CO$_2$ mixing ratio of 400 ppm,
and various flux rates of O$_2$ to the surface after 1 billion model years.}
\label{o2flux}
\end{figure}

The dependence of key atmospheric parameters on the O$_2$ flux rate to the surface
is shown in Figure \ref{o2flux}. Quantities are given after 1 billion model years.
Characteristics for an O$_2$ flux of $10^9$ molec.$\mathrm{cm}^{-2}\mathrm{s}^{-1}$ are very
similar to the results without any O$_2$ flux to the surface. The tropopause is found at a
temperature of $\sim$220 K, leading to a stratospheric water vapor mixing ratio of less than
200 ppm. Molecular oxygen has become the major constituent of the atmosphere, exceeding nitrogen
by a factor of 6. Maximum ozone levels around 20 ppb are found in the middle atmosphere.
Increasing the O$_2$ flux to $3-5 \cdot 10^9$ molec.$\mathrm{cm}^{-2}\mathrm{s}^{-1}$
brings the system into a regime where it approaches an equilibrium state within
1 billion model years. Tropopause temperatures are between 220 and 230 K, with corresponding
stratospheric water vapor mixing ratios. Oxygen mixing ratios are lower and in the case
with a flux of $5 \cdot 10^9$ molec.$\mathrm{cm}^{-2}\mathrm{s}^{-1}$ oxygen is not the main
atmospheric constituent anymore. Maximum ozone mixing ratios are still in the range of 10-20 ppm.
With an oxygen flux of $10^{10}$ molec.$\mathrm{cm}^{-2}\mathrm{s}^{-1}$ the behavior changes significantly.
The tropopause temperature stays above 230 K and the stratosphere stays much wetter, with
water vapor mixing ratios reaching equilibrium around 0.3\%. O$_2$ builds up only to a mixing ratio of 20\%.
The wet atmosphere inhibits the buildup of large ozone mixing ratios as ozone destruction by the catalytic
odd hydrogen cycle is increased, leading to maximum O$_3$ values of only 2 ppm. Hence a flux of
$10^{10}$ molec. O$_2$ $\mathrm{cm}^{-2}\mathrm{s}^{-1}$ is too strong to allow the buildup of abiotic oxygen
to a magnitude that would dry out the middle atmosphere.

\begin{figure}
\plotone{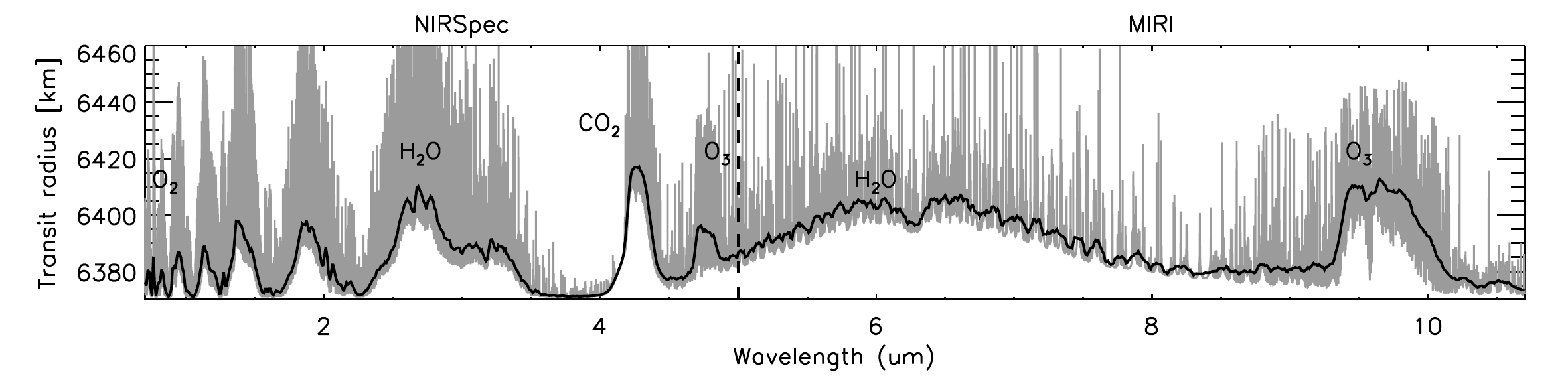}
\plotone{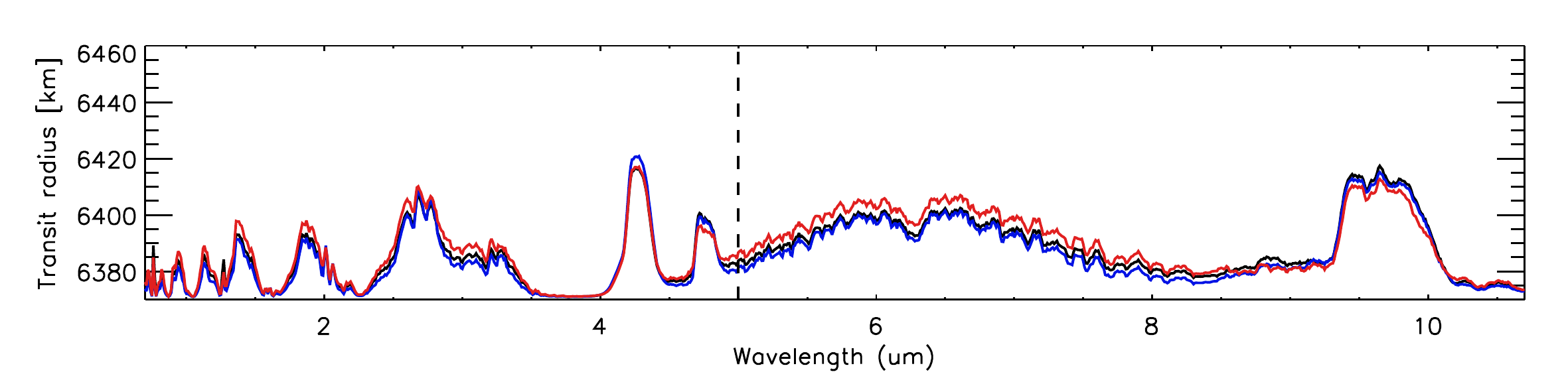}
\plotone{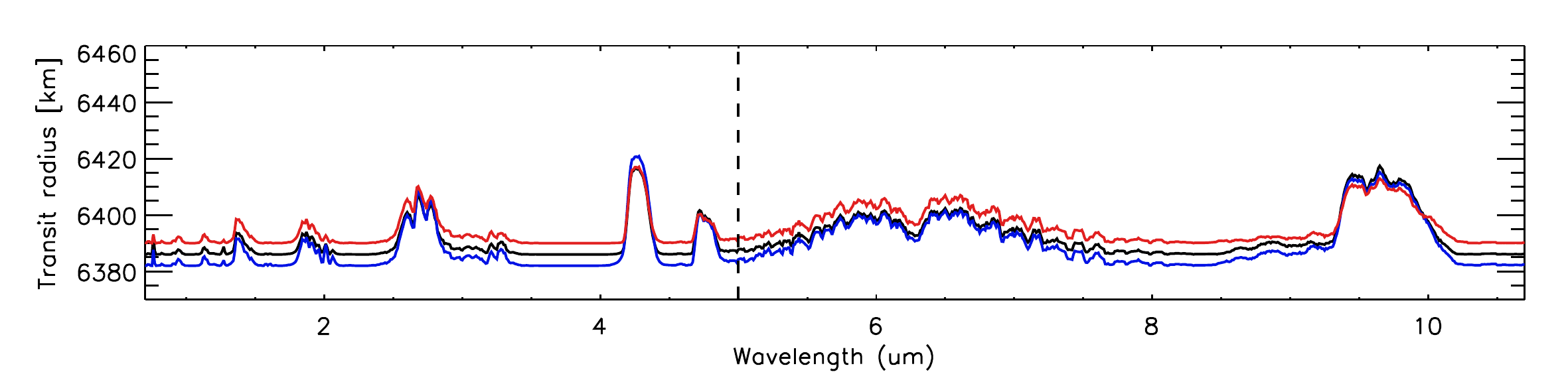}
\caption{Top: Spectrally resolved apparent radius of a transiting exoplanet 
with the atmospheric structure and composition based on Figure \ref{n202o2flux}
after 1 billion model years.
The gray lines give the extinction spectrum at high spectral resolution, 
while the black line shows the spectrum convolved to a resolving power of 100.
Center: A comparison of the low resolution transit spectrum from the top panel (red)
with a spectrum based on the scenario with Earth's inventory of N$_2$ and 21\% O$_2$
from Figure \ref{n210o2bio} (blue),
and with the spectrum from Figure \ref{n202spec} after 1 billion model years (black).
The bottom panel shows the transit spectra for the same atmospheric conditions
assuming extended cloud cover in the troposphere.}
\label{n210o2bioN202o2fluxspec}
\end{figure}

Figure \ref{n210o2bioN202o2fluxspec} shows comparisons of transit spectra
for the scenario without O$_2$ flux to the surface after 1 billion model years from Figure \ref{n202},
for the equilibrium scenario with an O$_2$ flux of $3 \cdot 10^9$ molec.$\mathrm{cm}^{-2}\mathrm{s}^{-1}$
after 1 billion model years from Figure \ref{n202o2flux},
and for the scenario with the Earth-like inventory of N$_2$ and O$_2$ from Figure \ref{n210o2bio}.
The striking feature of this comparison is that the transit signatures of ozone in the 9.6 $\mu$m band are
very dominant in the spectrum but with heights of 44-46 km they are virtually indistinguishable.
The oxygen A-band at 0.76 $\mu$m is with a transit height of 12-13 km
of comparable magnitude in the scenarios with O$_2$ flux to the surface and with biogenic oxygen,
and somewhat larger in the scenario without O$_2$ flux to the surface, where it has a transit height of 18 km.
The spectral features of water vapor are very similar and would likely be indistinguishable
in the scenarios without O$_2$ flux to the surface and with biogenic oxygen. The wetter middle atmosphere
of the scenario with O$_2$ flux to the surface causes a somewhat larger water signal, which exceeds
the transit radii of the other scenarios by about 5 km in the 6.5 $\mu$m region.

If we consider extended cloud cover in the troposphere (bottom panel of Figure \ref{n210o2bioN202o2fluxspec})
we notice that the most distinctive feature between these scenarios is the altitude at which the atmosphere
becomes opaque due to cloud cover. However, as the planetary radii are unlikely to be know precisely in
transit observations of exoplanets it would be hard to exploit this feature to distinguish between scenarios.
This is particularly problematic with water vapor estimates that would likely rely on differences in
transit radii in and out of water bands (e.g. at 2.7 $\mu$m vs. 3.5 $\mu$m or at 6.5 $\mu$m vs. 8 $\mu$m).
The higher tropopause in the case of the wetter middle atmosphere makes this difference more comparable
with the other scenarios. This can lead to erroneous interpretations of water vapor in transit spectra if this
effect is not taken into account through modeling.

\section{Summary and Conclusions}
\label{conc}

We have used a photochemical model coupled to a 1D radiative-convective equilibrium model to self-consistently model
the structure and composition in atmospheres with N$_2$, CO$_2$ and H$_2$O as the main constituents in a moist convective regime.
We find that in atmospheres with an N$_2$ inventory of order 20\% of Earth's N$_2$ inventory,
water vapor mixing ratios in the middle atmosphere can be over two orders of magnitude higher
compared to atmospheres with an Earth-like N$_2$ inventory. Without a strong surface sink,
oxygen builds up rapidly in the atmospheres of such planets. 

For an Earth-like planet with 20\% of Earth's N$_2$ inventory and low CO$_2$ mixing ratio around a Sun-like star,
the O$_2$ mixing ratio starts to exceed the N$_2$ mixing ratio after $\sim$10 million years.
Over the course of a billion years oxygen builds up to the extent that the surface pressure exceeds 1 bar.
The oxygen serves as a non-condensable that lowers the atmospheric cold-trap temperature by $\sim$15 K.
This restricts water vapor transport to the upper atmosphere and reduces the upper atmospheric
water vapor mixing ratio by over two orders of magnitude. The oxygen amount sustains an ozone
volume mixing ratio of order 30 ppm in the stratosphere. The atmospheric temperature structure
as well as the amounts and distributions of water vapor and ozone in these atmospheres are comparable
to the atmospheric conditions found on Earth.

Tropopause temperatures tend to decrease with an increase in N$_2$ inventory
and tend to increase with an increase in CO$_2$ mixing ratio. They exceed 240 K for low N$_2$ inventories
and CO$_2$ mixing ratios of order 1\% at 10000 model years.
All scenarios without O$_2$ flux to the surface considered in this work
ended up with stratospheric water vapor mixing ratios below 0.1\% and maximum ozone mixing ratios
between 10 and 30 ppm after 1 billion model years. In cases with oxidation of the surface,
as steady state is approached after 1 billion model years for fluxes of 
$3 \cdot 10^9$ molec. O$_2$ $\mathrm{cm}^{-2}\mathrm{s}^{-1}$ or greater, 
where the oxygen production through water vapor photolysis is balanced by the surface sink.
Such a planet could sustain a moderately moist middle atmosphere with
significant amounts of oxygen and ozone over extended periods of time.

We have used a radiative transfer model to study the spectroscopic fingerprint of these atmospheres
in transit observations with reference to the capabilities of JWST.
We find that abiotically produced oxygen can produce ozone amounts comparable to the ones found in Earth's atmosphere,
and spectral oxygen and ozone features be nearly identical to oxygen and ozone signatures expected from the biogenic oxygen
on Earth in a transit observation.
In addition, the abiotic oxygen dries out the stratosphere to the extent
that stratospheric water vapor mixing ratios can be in a comparable range as on Earth.
It has been suggested that oxygen abiotically produced by the photolysis of water vapor
can be recognized because it would be accompanied by high levels of middle atmospheric water vapor
in spectroscopic observations (e.g. \citet{meadows17}).
Our results suggest that this is not necessarily the case.
We find that the use of oxygen or ozone as a bioindicator is at best problematic,
given the possibilities for their entirely abiotic production described here.
Their interpretation as bioindicators will likely involve the use of atmospheric models
together with considerations concerning the plausibility of planetary parameters used as input to these models.

\section*{Acknowledgments}
This work was supported by the Jet Propulsion Laboratory Exoplanet Science Initiative program.
Work at the Jet Propulsion Laboratory, California Institute of Technology,
is performed under a contract with the National Aeronautics and Space Administration.
Copyright 2018 California Institute of Technology. U.S. Government sponsorship acknowledged.

\bibliographystyle{aasjournal}
\bibliography{abiot}




\end{document}